\documentclass[11pt]{article}

\usepackage{amsfonts,amsmath,amssymb,graphics}
\usepackage{multicol}
\usepackage{array}
\usepackage{arydshln} 
\usepackage{float}
\usepackage{subcaption}

\newcolumntype{L}[1]{>{\raggedright\let\newline\\\arraybackslash\hspace{0pt}}m{#1}}
\newcolumntype{C}[1]{>{\centering\let\newline\\\arraybackslash\hspace{0pt}}m{#1}}
\newcolumntype{R}[1]{>{\raggedleft\let\newline\\\arraybackslash\hspace{0pt}}m{#1}}
\newcommand{\notthis}[1]{}

\newcommand{\pr}{\mathbf{P}}
\newcommand{\ex}{\mathbf{E}}

\newcommand{\indic}[1]{\mathbf{1}(#1)}

\newcommand{\ISDARPV}{\widetilde{\Pi}}
\newcommand{\extra}{A}
\newcommand{\recov}{\Re}

\newcommand{\mma}{\mathfrak{B}}
\newcommand{\sbar}{\overline{s}}

\newcommand{\inv}{^{-1}}
\newcommand{\half}{\frac{1}{2}}
\newcommand{\pbar}{\overline{p}}

\newcommand{\cdl}{\,|\,}

\begin{document}

\title{\bf A CDS Option Miscellany}

\author{Richard J. Martin}

\maketitle

\begin{abstract}

CDS options allow investors to express a view on spread volatility and obtain a wider range of payoffs than are possible with vanilla CDS. The authors give a detailed exposition of different types of single-name CDS option, including options with upfront protection payment, recovery options and recovery swaps, and also presents a new formula for the index option. The emphasis is on using the Black'76 formula where possible and ensuring consistency within asset classes. In the framework shown here the `Armageddon event' does not require special attention.

\noindent First version: Dec 2009

\noindent Second version: Jul 2017 (index option section updated)

\noindent Third version: Apr 2019 (minor corrections)

\noindent This version: Dec 2025 (index option section updated, some simplifications, and new RPV01 section in \S1)

\end{abstract}


\section*{Introduction}

CDS options allow investors to express a view on spread volatility and give a wider range of payoffs than are possible with vanilla CDS. Index options are commonly traded: the reader is referred to \cite{Doctor07} for a discussion of trading strategies and hedging of index options. The market in the single-name options is not well developed, but the options are embedded in a number of structured products, for example cancellable single-name protection. With the standardisation of the single-name CDS and the move to central clearing, and the greater emphasis placed on hedging and CVA, it is possible that single-name CDS options will become more relevant, in the same way that equity options have been around for years. Accordingly, their definition and valuation merits study.

The valuation of single-name options with running spread was originally performed in the Black'76 framework some years ago \cite{Hull03,Schoenbucher03,Schoenbucher04}. In this article we briefly review it and develop the subject to point out what happens when the strike is part-upfront, by which we mean that on exercise the CDS protection is to be paid wholly or partly upfront rather than all-running---which is how contracts now trade\footnote{i.e. with fixed coupon, so that when trades are closed out, they net off with IO strip left over. This `Big Bang' was necessary for the introduction of centralised clearing.}. Upfronts introduce complications. The Black'76 formula is not directly applicable because on exercise the option is not being paid for in units of risky annuity, but instead in a mixture of cash and risky annuity. We could simply start with a hazard rate model formulation, and build everything up from first principles. This is the approach taken by more recent authors such as \cite{BenAmeur06,Brigo08a}.
However, we would lose equivalence with Black76 in the all-running case, even for a lognormal hazard rate model. (Lognormal hazard rate dynamics do not exactly correspond to lognormal credit spreads. Also, in the all-running case the spread is assumed for mathematical convenience to be lognormal in the survival measure with the risky PV01 as numeraire, not the money market account.) We wish to keep consistency with Black'76 as far as possible as that is the standard model used on trading desks, even if Brownian spread dynamics are `clearly not right'.
Another complication of upfronts is that the no-knockout option\footnote{Knockout (`KO') means that the option expires worthless if a credit event occurs before the option expiry; no-knockout (`NKO') means that it does not and can therefore be exercised into the defaulted name.} contains an embedded option on realised recovery, for which pricing models need to be developed. We deal with this too.

CDS indices are different from single-name options in that they trade no-knockout: the payoff arising from defaults in the life of the option is referred to as front-end protection (FEP). It is pointed out by Pedersen \cite{Pedersen03} that one cannot simply add the FEP to a knockout payer option to get the no-knockout price: the exercise decision depends on the accrued loss. He addresses this problem by introducing a loss-adjusted spread which he then models as lognormal. But Brigo (\cite{Brigo09a} \& references therein) points out that spread-based formulae have a problem: to derive them rigorously requires changing numeraire to the associated PV01. In the `Armageddon' event of all names defaulting, the numeraire collapses to zero and the spread becomes undefined, so this event has to be conditioned against and treated separately. Brigo \& Morini seek to estimate the probability of collapse using the tranche market, but this market has hardly been a paragon of efficiency, and they use the Gaussian copula model which did not even calibrate during the crisis: also, what is one to do with an index on which no tranches are traded (e.g.\ ITX Financials, or ITX.XO in times gone by)? 

Most academic literature, including \cite{Brigo09a,Rutkowski08,Sveder15,Herbertsson19}, fails to treat the payoff of a CDS index option correctly. A CDS index is not a portfolio of par single name CDS: it trades upfront with a fixed coupon and in the options the form of the strike leg payment needs careful attention. 
In fact, Rutkowski \& Armstrong \cite[Eq.42]{Rutkowski08} start by writing down the payoff correctly, then remark that it is difficult to handle analytically (which is true) and that it is ``internally inconsistent'' (as we explain later, it is not) and then proceed to model something else instead. 
With the exception of that paper, it appears that academic authors are unaware of how credit instruments trade, and so they end up (inventing and then) solving rather uninteresting problems. 
For example, Herbertsson's assertion \cite{Herbertsson19} that lognormal distributions are not ideal is, in common with other areas of finance, correct, but introducing significant extra complexity into a model and making pronouncements about real-world pricing seems to be of dubious value when the equation for the instrument's payoff \cite[Eq.(2.3.1)]{Herbertsson19} is incorrect at the outset.
Of the practitioner literature, Pedersen \cite{Pedersen03} and O'Kane \cite{OKane08} get it right, though O'Kane's exposition is clearer.
A CDS index option is physically settled, i.e.\ it exercises into an index CDS contract, even if the  resulting contract is immediately closed out. In other words it is not an option on a \emph{spread} but rather an option on a \emph{contract} (PV), and more precisely it is the difference of two `Bloomberg CDSW' calculations, one for the spot and one for the strike. {\bf Therefore the starting-point must be to figure out exactly how that calculation works;} judicious approximations can then be made but they have to respect the underlying nature of the problem.
Finally, returning to the `Armageddon event': this is not a difficulty in our framework, and so one does not need to estimate its probability, which is just as well as such a procedure would be little better than guesswork.

\vspace{3mm}
\noindent {\bf Notation.} RPV01 means the PV of a risky annuity (IO strip) on the issuer; the symbol for its value on date $t$ will be $V^1_t(T_1,T_2)$, indicating the dates $T_1,T_2$ between which the annuity is to be paid. The value of the annuity collapses to 0 on default. By the `default-PV' of a CDS, $V^D_t(T_1,T_2)$, we mean the PV of the protection leg. As usual $B$ denotes the riskfree discount factor (riskfree zero-coupon bond price), $\mma_t$ the rolled-up money market account ($\mma_0=1$), and for single issuers $\tau$ is the default time and $Q(t)$ the survival probability. $C_+,C_-$ generically denote a call or put on some underlying, the interpretation being obvious from the context.

\section{How CDS trade, and a new RPV01 formula}

As already stated, we provide a reminder that CDS are not par instruments, but trade with a fixed coupon on standardised payment dates with an upfront being taken to offset the difference between the traded spread and the coupon.
The PV of this difference is naturally the spread difference multiplied by `the' risky PV01. The trouble with this is that although it is not conceptually difficult, knowledge of the RPV01 requires agreement of the entire CDS curve up to the traded tenor, and this is a practical impossibility. The standardisation from 2009 enforced a flat curve assumption, so that the ISDA RPV01, which we are denoting $\ISDARPV$, is a sort of `made-up number', essentially derived by applying a constant hazard rate $\tilde{\lambda}=\tilde{s}/(1-\recov)$ across all maturities, with $\tilde{s}$ denoting the traded spread and $\recov$ the `valuation recovery'. (Or that is what should have been done, but the standardised method relies on using flat spreads and stripping the curve. This can fail when spreads are very high.) 

As the par curve is generally upward-sloping except for distressed credits, the ISDA RPV01 is generally lower than the true one. A consequence of this is that the par spread and the traded spread differ, and the traded spread is coupon-dependent. That the traded spread is turned into an upfront using an `incorrect' RPV01 is not of itself a problem. Although it might seem that across the curve, inconsistent assumptions are made (i.e.\ a 1y traded spread is turned into an upfront using a flat curve at the 1y rate whereas a 5y traded spread is turned into its upfront using a flat curve at the 5y rate), this simply manifests itself in the traded spread differing from the par spread.

The `valuation recovery' also causes confusion. Its \emph{only} use is to turn a quoted spread into an upfront for settlement purposes. The mystical 40\% number (or, strangely, 30\% for US HY indices, though not the single names) has very limited relevance. Names that trade upfront do not, therefore, make any reference to it. It follows that the correct way to strip CDS curves for use in other models is to turn all spreads into upfronts, after which \emph{any recovery can be used to impute the survival curve}. Further, the recovery swap market, inasmuch as it exists, can trade where it likes (within reason) and although for iTraxx Main it is typically around 40\%, there is no arbitrage to be had if it trades at some different level.

This brings us to an important matter: calculating $\ISDARPV$. At some basic level this cannot be difficult, and just requires the relevant OIS/ESTR/SOFR curve and $\tilde{\lambda}$. But the reality is that this is quite a painful calculation to do `on the fly' (CDSW does it, but this is not much use outside the confines of Bloomberg), and far worse to do historically, which is necessary for computing the PL of trades or strategies where, as is usually the case, the historical information is the credit spread.

In the simple case where the riskfree zero curve is flat, no problem presents itself, and the RPV01 in continuous format is the oft-quoted formula
\begin{equation}
\ISDARPV(T) \stackrel{?}{=} \frac{1-e^{-(r+\lambda)T}}{r+\lambda}.
\label{eq:RPVsimple}
\end{equation}
The problem is that if the riskfree curve is not flat, discrepancies creep in, especially for longer-dated contracts. Really, one wants a formula that agrees with the swap PV01 when $\lambda=0$,
\begin{equation}
\Pi^\circ(T) = \frac{1-e^{-r_Z T}}{r_S} ,
\label{eq:Pi0}
\end{equation}
where $r_Z$ is the zero rate (usually quoted continuously-compounded, Act/365) and $r_S$ the swap rate for the appropriate cash-flow schedule (Q Act/360, in context). The rate curve in question is the OIS/SOFR/ESTR curve in the correct currency\footnote{Bloomberg: S0042Z for USD, S0514Z for EUR.}.
These rates will be available at various tenors, and historically too, and one either interpolates them and then uses the resulting $\Pi^\circ$ or does two $\Pi^\circ$ calculations and then interpolates the result: the whys and wherefores of which is better will not concern us here, though we suggest that the former is easier. 

The next formula is something of an enigma, because it is not clear why it works as well as it does. It is best described as the `riskification formula' and states:
\begin{equation}
\boxed{
\ISDARPV(s;T) \approx \Pi^\circ(T) \cdot \frac{1+x/2}{1+x+x^2/2}, \qquad x = \lambda \cdot \Pi^\circ(T)
}
\label{eq:newRPV}
\end{equation}
with $\lambda=s/(1-\recov)$ as usual.

To an extent this result is reminiscent of (\ref{eq:RPVsimple}): if we take that and divide by $T$ we get $(e^y-1)/ye^y$, $y=(r+\lambda)T$, after which Taylor series expansion of both exponentials gives something similar. However, they are still not the same because we have defined $x$ not as $\lambda T$ but as $\lambda \Pi^\circ(T)$---the reason being that as $\lambda\to\infty$ we must have $s\ISDARPV(s;T)\to 1-\recov$, which is equivalent to $\ISDARPV(s;T)\sim \lambda\inv$, and (\ref{eq:newRPV}) does this by design. Incidentally this is a good advertisement for Pad\'e approximants: rational functions allow behaviour at $0$ and $\infty$ to be captured in the same formula.

For moderate spreads (0 to a few hundred bp) this formula gives excellent results, usually correct to 2--3 d.p.\ which for practical work is enough. For high spreads (1000bp or more) the accuracy declines, but then these have little economic relevance as the resulting instruments are quoted upfront. Once the spread gets very high, the asymptotic properties ensure that accuracy improves once more.

An obvious question is why the Taylor series development, alluded to earlier, is truncated after the quadratic term. The reason is that one often needs to solve the inverse problem, i.e.\ find the traded spread given the upfront: in other words, solve
\begin{equation}
u = (s-c)\ISDARPV(s;T)
\end{equation}
for $s$, given $\Pi^\circ(T)$, $c$, $\recov$ and $u$. By (\ref{eq:newRPV}),
\[
\left( x - \frac{c\Pi^\circ(T)}{1-\recov} \right) \frac{1+x/2}{1+x+x^2/2} = \frac{u}{1-\recov} 
\]
which rearranges to
\begin{equation}
\left( 1-\frac{u}{1-\recov} \right) \frac{x^2}{2}
+ \left( 1- \frac{u+c\Pi^\circ(T)/2}{1-\recov} \right) x
- \frac{u+c\Pi^\circ(T)}{1-\recov} = 0 .
\label{eq:invRPV}
\end{equation}
This can be solved instantly, and because $-c\Pi^\circ(T) < u < 1-\recov$ (to avoid arbitrage), there is a unique positive solution in $x$, after which $s=x(1-\recov)/\Pi^\circ(T)$.
Higher-order approximations might offer an improvement, but would require the solution of polynomial equations of higher degree, either directly or by iterative treatment, i.e.\ guess $s$, compute a new value of $s$ given by $u/\ISDARPV(s,T)+c$, and repeat until convergence.

Given that the existing method is already an approximation, or put less politely a `made-up calculation', there seems to be nothing wrong with using (\ref{eq:newRPV}), in conjunction with (\ref{eq:Pi0}), as the \emph{exact} ISDA RPV01 formula, with (\ref{eq:invRPV}) providing the upfront-to-spread conversion. This is something that the `authorities' might be well advised to consider.

\newpage
\section{Single-name options}

\subsection{Knockout, running}

In the old days, single-name options were assumed to trade knockout and exercise into  protection paid as a running premium.
The payer price $C_+$ (the receiver price will be analogous and we omit it throughout) is
\begin{eqnarray*}
C_+^\textrm{ko,r} &=& \ex_0 \left[\frac{1}{\mma_{t_E}} \big(V^D_{t_E}(t_E,T)-s_KV^1_{t_E}(t_E,T)\big)^+  \indic{\tau>t_E}  \right] \\
&=& \ex_0 \! \left[(\sbar_{t_E}-s_K)^+ V^1_{t_E}(t_E,T) \right]
\end{eqnarray*}
where $\sbar$ denotes the par spread, $s_K$ the strike spread,  $V^1_t$ the RPV01, $t_E$ the expiry date of the option, and $T$ the maturity date of the underlying swap (today=0). Now $V_t^1$  is a collapsing numeraire as it is zero after default, so we pass to the survival measure \cite{Schoenbucher04} $\ex^*$ defined as  
\[
\ex^* [Z] \equiv \frac{\ex[\mma_{t_E}\inv Z V^1(t_E,T)\indic{\tau>t_E}]}{\ex[\mma_{t_E}\inv V^1(t_E,T)\indic{\tau>t_E}]};
\]
note that in the denominator the survival indicator is unnecessary as  $V^1_t=0$ if default occurs before time $t$.
Then
\[
C_+^\textrm{ko,r} = \ex^*_0 \left[(\sbar_{t_E}-s_K)^+\right] V^1_0(t_E,T).
\]
Under $\ex^*_0$, $\sbar_t$ is assumed to be lognormal, and then the expectation evaluates using Black'76 to get:
\begin{eqnarray}
& C_+^\textrm{ko,r} = \big(\sbar_F \Phi(d_+) - s_K \Phi(d_-) \big) V^1_0(t_E,T) & \label{eq:pkor2} \\
& d_\pm = \displaystyle \frac{\ln (\sbar_F/s_K) \pm \half \sigma^2 t_E}{\sigma \sqrt{t_E}}. & \nonumber
\end{eqnarray}
Note that  $\sbar_F$ (the knockout forward spread) is equal to $V^D_0(t_E,T)/V^1_0(t_E,T)$ until default, at which point it becomes undefined. When performing any statistical analysis on the credit spread, for example to estimate realised volatility, one is naturally observing it prior to default anyway, so the survival measure is the natural frame of reference.

\subsection{No-knockout, running}

A no-knockout option remains valid after default. By splitting it into a survival-contingent and a default-contingent part, one sees that the NKO payer must be worth the sum of the KO payer option and the FEP, so
\[
C_+^\textrm{nko,r} = C_+^\textrm{ko,r} + V^D_0(0,t_E).
\]
A NKO receiver option is worth no more than a knockout receiver option because a receiver option is never exercised into a defaulted credit: one would  receive no running spread but pay par minus recovery. However, if the premium leg on exercise is partly paid upfront, the position is more difficult, as we are about to see.

\subsection{Knockout, upfront+running}

If the premium on exercise is to be paid partly (or wholly) upfront, the expression for the payer option is now
\begin{equation}
C_+^\textrm{ko,ur} = \ex_0\! \left[\frac{1}{\mma_{t_E}} \big(V^D_{t_E}(t_E,T)-u_K - s_KV^1_{t_E}(t_E,T)\big)^+  \indic{\tau>t_E}  \right]
\label{eq:pkour1}
\end{equation}
where $u_K$ is the upfront part of the strike and $s_K$ is the running part (in practice this is likely to be 100bp or 500bp). The RPV01 in the strike leg is as before and at maturity it depends on $\sbar_{t_E}$.

If we change numeraire in the same way as before and try to use
\[
u_K + s_KV^1_{t}(t_E,T)
\]
as numeraire, we end up with a valid option-pricing formula, but there are some problems. The main one is that it is inconsistent with the previous setup to the extent that one cannot use the same volatility for both.
Of the two PV ratios
\[
\left(\frac{V^D_0(t_E,T)}{s_KV^1_0(t_E,T)}\right) \mbox{ and } \left(\frac{V^D_0(t_E,T)}{u_K+s_KV^1_0(t_E,T)}\right),
\]
(conditionally on no default) the former has higher volatility, for when the default-PV goes up, the RPV01 goes down. We therefore expect a CDS option with a strike quoted partly upfront to be worth \emph{less} than that of an all-running one,  all other things being equal. Another objection is that the second quantity is bounded, whereas the first is not, as $s_t\to\infty$, so a lognormal distribution is arguably inappropriate; however this should only be an objection for very high strikes.

To ensure consistency with (\ref{eq:pkor2}) we therefore need to stick with lognormal spread dynamics in the $V^1_t$-measure\footnote{i.e.\ the survival measure with RPV01 as numeraire.} (which we are calling $\pr^*$).  However, the direct computation of the option price (\ref{eq:pkour1}) has to be done in the $\mma_t$-measure $\pr$. A change of numeraire is required.
Under $\pr^*$ the spread is lognormal, so the expectation of some function $G$ of the spread at time $t$ is
\[
\ex^*_0[G(s_t)] = \int_{-\infty}^\infty G\big(\sbar_t(z)\big) \phi(z) \,dz, \qquad \sbar_t(z)=\sbar_F e^{\sigma z \sqrt{t}-\half \sigma^2 t}
\]
with $\phi(z)=\frac{1}{\sqrt{2\pi}}\exp(-\half z^2)$ and $\sbar_F$ denoting the forward spread. (Clearly this prescription allows the model to be embedded in a Markovian spread model, which is necessary for term structure models but not for the one-period models here.)
Next, note that for any random variable $Z$,
\begin{equation}
\ex\!\left[\frac{Z}{\mma_t}\indic{\tau>t}\right] \equiv \ex^*\!\left[\frac{Z}{V^1_t}\right] \ex\!\left[ \frac{V^1_t}{\mma_t} \indic{\tau>t}\right].
\label{eq:chgmeas}
\end{equation}
We need to link the RPV01 to the spread, i.e.\ write $V^1$ as a function of $\sbar$, and to do this we use the standard RPV01 calculation $\ISDARPV$ based on a flat hazard rate curve and constant `assumed' (or marking) recovery rate $\recov$. For a flat riskfree curve this is
\begin{equation}
\ISDARPV(s;t,T) = \frac{1-e^{-[r+(1+\varepsilon)s/(1-\recov)](T-t)}}{r+(1+\varepsilon)s/(1-\recov)}
\label{eq:DV01flat}
\end{equation}
The parameter $\varepsilon$, which is small, will be explained presently.
First, set $Z\equiv1$ in (\ref{eq:chgmeas}) to obtain
\[
 \int_{-\infty}^\infty \frac{\phi(z)\,dz}{ \ISDARPV  \big(\sbar_{t_E}(z);t_E,T\big)  } = \frac{B^*(t_E)}{V^1_0(t_E,T)}
\]
in which the RHS is obtained from today's CDS curve and $B^*(t_E)$ is the risky discount factor.
The purpose of $\varepsilon$ is to ensure that this relation is satisfied exactly, and a unique $\varepsilon$ can always be found to do the job\footnote{For $\varepsilon\to\infty$, $\ISDARPV(s)\to0$; and for $\varepsilon=-1$, $\ISDARPV(s)$ is the riskfree PV01 which is as high as it can be.}.
Therefore we have matched the default and premium legs of today's CDS curve, by matching the forward rate and RPV01.
We now have
\begin{equation}
\ex_0\!\left[\frac{G(\sbar_{t_E})}{\mma_t}\indic{\tau>{t_E}}\right] =
\int_{-\infty}^\infty \big[G\big(\sbar_{t_E}(z);t_E,T\big) \big/ \ISDARPV \big(\sbar_{t_E}(z);t_E,T\big)\big] \phi(z)\, dz
\cdot
V^1_0(t_E,T)
\label{eq:chgmeas2}
\end{equation}
and can now calculate the option price numerically from (\ref{eq:chgmeas2}).
This will give consistency with the Black'76 formula, because if the option exercises into all-running protection then the integral (\ref{eq:chgmeas2}) gives the same result as (\ref{eq:pkor2}).

\subsection{Knockout, Standard American/European}

CDS are now quoted running but trade upfront with a fixed coupon $c$ say. This is very much like the previous section, but the RPV01 is slightly different. There is no standard convention for the options, and the payoff is best written:
\begin{equation}
C_+^\textrm{ko,ur} = \ex_0\! \left[\frac{1}{\mma_{t_E}} \big( (s^q_{t_E}-c) \ISDARPV_{t_E}(s^q_{t_E};t_E,T) - (s_K-c) \ISDARPV_{t_E}(s_?; t_E,T) \big)^+  \indic{\tau>t_E}  \right]
\label{eq:pkour2}
\end{equation}
where the notation $s^q$ indicates the quoted spread, as distinct from the par spread $\sbar$ [the two are related by $(s^q-c)\ISDARPV = (\sbar-c)V^1$]. The symbol $s_?$ in the strike RPV01 reflects uncertainty as to what spread to use: it could be the spot at expiry ($s^q_{t_E}$) or the strike spread ($s_K$). Whichever one is chosen, equation (\ref{eq:chgmeas2})  can be used to value the option via a numerical integral. In this way we ensure consistency between the old-style CDS and the new-style with different coupons.

\subsection{No-knockout, upfront+running}

When part of the CDS premium on exercise is to be paid upfront, the no-knockout receiver option has more value than the knockout.
This is because the holder of a receiver owns a call on the realised recovery $\recov$, with strike equal to 100\% minus the upfront. For example if the receiver is struck at 12\% plus 500bp running, and a default occurs with 92\% recovery, the option holder can exercise and make 4\%. Similarly the payer has an embedded put:
\begin{eqnarray*}
C_+^\textrm{nko,ur} &=& C_+^\textrm{ko,ur} + C_-(\recov;1-u_K) \\
C_-^\textrm{nko,ur} &=& C_-^\textrm{ko,ur} + C_+(\recov;1-u_K) .
\end{eqnarray*}
One therefore has the task of valuing the recovery option.

\subsection{Recovery swaps and options}

\begin{figure}[!h]
\centerline{\scalebox{0.7}{\includegraphics*{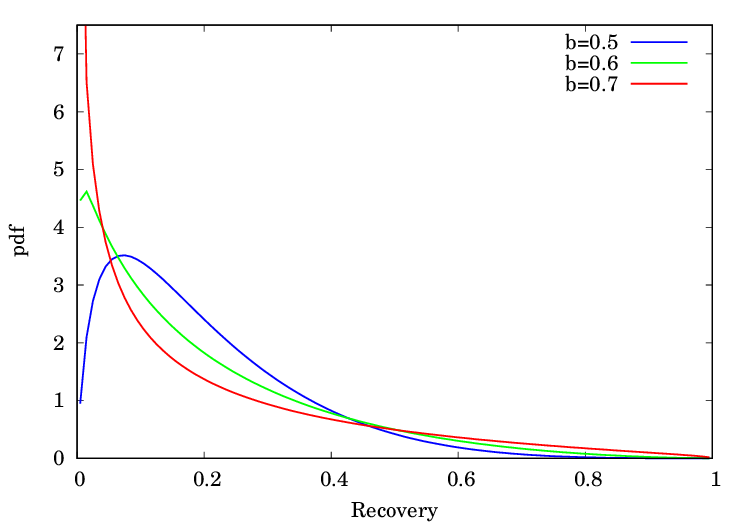}}}
\caption{Vasicek distributions with mean 0.2 and width parameter 0.5, 0.6, 0.7.}
\label{fig:f1}
\end{figure}

Recovery is obviously bounded to $[0,1]$ so we wish to choose an appropriate distributional assumption. Although a Beta distribution is often suggested, we suggest the use of a Vasicek distribution instead as these are a little more tractable, in the sense that for option pricing one only needs the bivariate Normal distribution as opposed to the incomplete Beta function. Also the use of the Vasicek distribution ties in neatly with probit modelling of recovery rates. The recovery $\recov$ is modelled as a transformation of a Normal variable, thus:
\[
\recov(Z) = \Phi\!\left(\frac{a+bZ}{\sqrt{1-b^2}}\right), \qquad Z\sim N(0,1)
\]
which has mean $\Phi(a)$ and variance $\Phi_2(a,a;b^2)-\Phi(a)^2$. (Here $\Phi_2(x,y;\rho)$ denotes as usual the cumulative bivariate Normal distribution.)  For small $b$ the standard deviation is roughly $\phi(a)b$, which follows from the tetrachoric series expansion of $\Phi_2$. This enables the distribution to be parametrised in terms of understandable numbers. Notice also that, one can easily model correlated recovery rates of several issuers, if desired, just by correlating their $Z$-variables through a multivariate Normal.
Figure~\ref{fig:f1} shows the density for mean $20\%$ ($a=-0.842$) and three different values of the width parameter $b$ (0.5,0.6,0.7). 

The payoffs of a recovery swap (or lock), call, and put are, respectively, $\recov-K$, $(\recov-K)^+$ and $(K-\recov)^+$, at expiry, where $K$ is the recovery strike. There is no payment if no default occurs.
Using the result
\[
\ex\! \left[\Phi\!\left(\frac{a+bZ}{\sqrt{1-b^2}}\right)
 \indic{Z>\xi} \right] = \Phi_2 (-\xi, a; b)
\]
we can obtain the call and put payoffs with strike $u$ as
\begin{eqnarray*}
\ex[(\recov-u)^+] &=& \Phi_2(a,c;b) - \Phi(c)u \\
\ex[(u-\recov)^+] &=& \Phi(-c)u - \Phi_2(a,-c;-b)  
\end{eqnarray*}
with $c=\big(a-\Phi^{-1}(u)\sqrt{1-b^2}\big)/b$. That put-call parity is satisfied is evident from elementary symmetries in $\Phi_2$.

The option values $C_\pm(\recov;u)$ are these expected payoffs multiplied by the difference between the risky and riskfree discount factors (i.e.\ $B(t_E)-B^*(t_E)$), because the cashflow occurs at time $t_E$ and is contingent upon default before option expiry. The recovery swap PV is obviously the difference between the strike and market recovery level, multiplied by the same factor.

In practice one obviously has to estimate the two parameters $a$ and $b$. Recovery locks do trade so there is some guide from the market as to where recovery is likely to be, but there is no information about uncertainty in recovery.
As an example, take the defaults in CDX.HY9 in 2008-09. The average recovery was 17.5\% but the dispersion was very wide: the highest was Tembec with 83\%, there were a couple at supposedly investment-grade levels (Quebecor 41.25\%, Lear 38.4\%), but the lowest were almost zero (Tribune 1.5\%, Idearc 1.75\%, Charter 2.375\%, Visteon 3\%, Abitibi 3.25\%, RHD 4.875\%). Thus without any knowledge of the credit one should probably use a high dispersion parameter ($b\sim 0.75$ to fit these points\footnote{By Kolmogorov-Smirnov.}). This phenomenon has persisted, and in recent times some credits (Europcar, Vue) recovered 100\% for technical reasons related to deliverability of the underlying bond, while others such as Casino Guichard, Atos, Rite Aid and Diamond Sports, had unsecured bonds that were pushed down so far in the capital structure by priming, and/or a worthless asset base, that recovery was next to zero. Recovery of 40\% is, therefore, something of an unusual phenomenon.
Of course, there is generally less uncertainty in the market over a particular traded credit, particularly at short horizon, and then the $b$ parameter is selected so as to represent the analysts' uncertainty.

One might ask whether it is necessary to `correlate recovery with default'. First one has to ask what this means. Importantly, the random variable $Z$ is only observed when default occurs, so one only wants its distribution conditional on default: in any other state of the world its value is irrelevant. Its distribution may be time-dependent (so that one has to use a different $a,b$ for different option maturities)---but use of different parameters for different maturities, most notably of course the Black-Scholes volatility, is standard practice.

For multivariate models, the question of correlation takes on a different form, and this is not relevant to the pricing of single-name options. One obviously wishes to correlate default events and there is evidence that when default rates \emph{en bloc} are higher, recoveries are lower \cite{Bruche08,Dobranszky08}. Incidentally the incorporation of this effect is also valuable in modelling of CDO tranches as it is a convenient device for pushing losses up into the senior tranches, a perennial difficulty for tranche modellers. Now, suppose that at a particular horizon we are to correlate defaults and recoveries for many credits. We want to ensure that we can consistently model correlated defaults and recoveries, having specified that the distribution of recovery (conditional, as said before, on default) is Vasicek. It turns out that this is possible without restricting the dependence structure, and this is discussed in the Appendix.

\subsection{Example}

We consider option pricing with the following parameters: vol 100\%, recovery 20\%, pricing date 09-Nov-09, option expiry 20-Mar-10, swap maturity 20-Dec-14. For no-knockout options we need the recovery rate volatility, and this is given via the `$b$' parameter which we take to be 0.6.

\begin{figure}[!h]
\centerline{\scalebox{0.6}{\includegraphics*{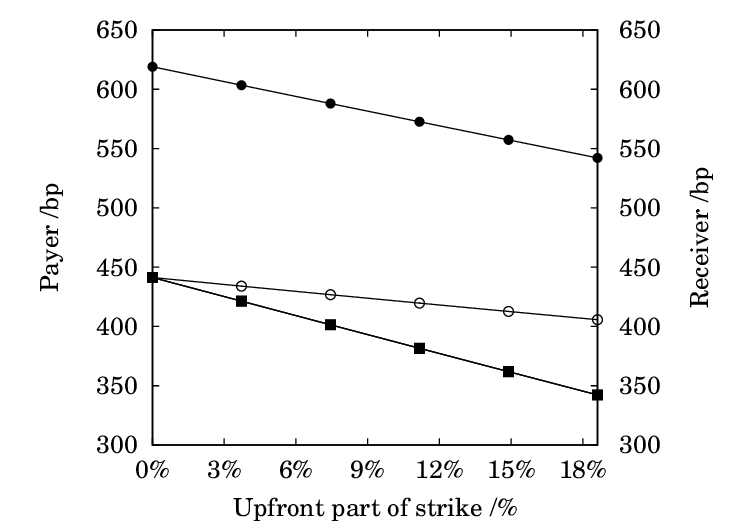}}}
\caption{Knockout and no-knockout option prices as a function of the upfront part of the strike; spot 500bp. Payers: {\Large$\circ$}=KO, {\Large$\bullet$}=NKO, both on LH axis. Receivers: {\scriptsize$\square$}=KO, {\scriptsize$\blacksquare$}=NKO, both on RH axis (they differ by $<0.1$bp). Running part of strike decreases linearly from 500bp to 0bp.}
\label{fig:f2}
\end{figure}

Figure~\ref{fig:f2} is for spot=500bp (forward-starting RPV01 3.723). The leftmost points on the graph are the payer and receiver premiums for all-running strike (500bp), which can be checked against the standard Black-Scholes price. The other points show the prices of options in which the strike is quoted partly upfront: the conversion is that 100bp running corresponds to 3.72\% upfront, and the all-upfront strike is 18.6\%.
As expected the option prices decrease as the proportion of the upfront increases. The KO payer should be 35bp cheaper for all-upfront strike than for all-running, and the receiver about 100bp cheaper. The NKO payer options are obviously more expensive than their KO counterparts, by an amount that decreases as the upfront part is raised. For the receiver the embedded recovery call option is too far out of the money to have noticeable value.

\clearpage

\section{Index options}

\subsection{General}

Index options are best treated as a separate asset class from single-name options because a literal model of an index option would require a model of correlated spreads and defaults between all the issuers.
Such an exercise (bottom-up approach) is not necessarily a bad idea, particularly if one wants to consider where the market for index options `should' be trading. But while this is potentially useful for strategising it is probably not useful for pricing. As likely as not, it will fail to match the market and then one has to work out which of the many parameters to adjust.
By convention they trade no-knockout, so that exercise is into the `original' index without removing any credits that have defaulted in the life of the option, and this is the case that we analyse here.
 As pointed out by O'Kane \cite{OKane08}\footnote{This is the most coherent treatment of CDS options we have seen and the reader is strongly advised to work through his numerical examples.}, the decision to exercise or not depends on the accrued loss in the pool, as well as the current spread and RPV01, so it is not possible to simply add the front-end protection to the payer price in the way that is done for single-name options.

Very few references treat the payoff correctly: it is not an option on a spread, but is instead physically settled, and its valuation is understood as the difference of a `spot' and a `strike' index contract, which means that to value it we need find the difference between two Bloomberg CDSW calculations. It follows, incidentally, that the most efficient way to calculate the payoff on expiry day is simply to run CDSW at each relevant strike; and as the number of active strikes is fairly small, this is not onerous. Nonetheless, the banks, to whom the task of doing this calculation usually falls, can still prove unequal to the task of producing the right number, especially when names have defaulted.

On exercise of an index payer, the following payments are made:
\begin{itemize}
\item
Option buyer receives par minus recovery on all defaulted names since the option was struck;
\item
Option buyer exercises into an index contract, at the prevailing market spread, on the remaining notional, i.e.\ the original notional multiplied by the proportion of names that are still undefaulted;
\item
Option buyer pays a cash amount equal to the strike spread minus the index coupon, multiplied by a PV01 defined by convention as $\ISDARPV_{t_E}(s_K;t_E,T)$, that is, the RPV01 assuming a flat curve at the \emph{strike} spread\footnote{Not the prevailing market spread: in passing we note that that would be undefined in the Armageddon event.}, on the \emph{full} notional.
\end{itemize}
It is also possible to exercise into the version of the index that was extant when the option was struck, and pay the difference from exercising the defaulted names: that gives the same result, as it must.
Symbolically, we have that the value of an index payer is
\[
C^\textrm{+idx}_t = \ex_t \left[\frac{\mma_{t}}{\mma_{t_E}} \left(L_{t_E}+\frac{N_{t_E}}{N_{00}}(s_{t_E}-c) \ISDARPV_{t_E}(s_{t_E};t_E,T)-\frac{N_0}{N_{00}}(s_K-c)\ISDARPV_{t_E}(s_K;t_E,T)\right)^+   \right]
\]
Here $\ISDARPV_t(s;T_1,T_2)$ is the `flat-hazard-rate' PV01 calculation alluded to above. It is simply a formal calculation to get from a spread to an upfront for use in settlement, and does not, for example, depend on the number of defaults in the index (though it does depend on an assumed recovery rate which is fixed by market convention), and nor is it a tradable asset as such. It is worth noting that the quantity
\[
\frac{N}{N_{00}} (s-c) \ISDARPV(s;t,T)
\]
is obtainable immediately from the Bloomberg CDSW screen, because it is the upfront (settlement amount) of an index CDS contract; so a market participant can easily calculate it. It is therefore consistent with CDS index option pricing methods.
The quantity $L_t$ is the accrued loss through defaults\footnote{Assuming interest is paid from the auction payout, but this is a minor matter. Notice, by the way, that the formula for $L_t$ after Eq.(42) in \cite{Rutkowski08} is incorrect, as the recoveries of the defaulted names may be different.} from time 0 to time $t$ and $N_t$ is the number of survived names at time $t$. 
The option is assumed to have been struck at time 0, and $N_{00}$ denotes the number of names in the original index: for example we might have $N_{00}=125$, $N_0=124$ and $N_t=122$, meaning that the index originally had 125 names, one default occurred, then the option was written, then another two occurred, and now it is time to value the option.
Also, $s_t$ is the spread of the version of the index current at time $t$ (i.e.\ with names that defaulted before time $t$ removed; $L_t$ keeps track of the defaulted ones).

A minor point worth ironing out is what happens when the strike spread is made arbitrarily high.
According to O'Kane, the payer option should become worthless in that limit, on the grounds that one is paying an `infinite spread' on exercise, whereas the simple procedure of taking a knockout payer Black'76 style and adding on the front-end protection causes the payer value to tend to the value of the front-end protection. In fact, both are wrong, because the exercise \emph{premium} does not tend to infinity: its limit is, as can be easily verified,
\[
\lim_{s_K\to\infty} (s_K-c)\ISDARPV_t(s_K;t_E,T) = 1-\recov
\]
where $\recov$ is the index recovery used in the RPV01 calculation. The `infinite spread' argument is a fairly common error, made by those who forget that to turn the spread into a monetary amount requires multiplication by an RPV01 which is, of course, spread-dependent.
Although older implementations of the Bloomberg CDSO page were faulty in this respect, the current one seems correct: as $s_K\to\infty$ the receiver price tends, more or less, to $(1-\recov)$.

\subsection{The forward spread}

In other asset classes, notably FX, there is a very liquid forward market and the forward is to be regarded as an input to the pricing model. Although that is not at all the case with CDS index options, the idea of the forward spread has carried over,  possibly as a result of earlier work on single-name options in which it does play a role.
Some implementations do use it as an input, but one can proceed without it, as follows.

The no-knockout forward spread $s_F$ is defined to be the fair spread for paying CDS premium over $[t_E,T]$ and receiving protection over $[0,T]$. If we follow the usual argument that buying a CDS forward and selling today should have zero PV, then the forward spread is given by the solution to
\begin{equation}
(s_F - c) \ISDARPV_0 (s_F; t_E,T) = (s_0 - c) \ISDARPV_0 (s_0; 0, T) + c \Pi^*_0(0,t_E)
\label{eq:sfwd}
\end{equation}
with $\Pi^*$ the true risky PV01(\footnote{As opposed to the ISDA calculation; $\Pi^*$ uses the whole CDS curve rather than the flat-hazard-rate assumption.}).

If assume $\ISDARPV(s_F)\approx \ISDARPV(s_0)$ and $\Pi^*\approx\ISDARPV$ (ignoring the index coupon is another way of getting there) the forward spread is approximately
\[
s_F \approx s_0\frac{\ISDARPV_0(s_0;0,T)}{\ISDARPV_0(s_0;t_E,T)}.
\]
The forward is therefore higher than the spot. The main reason for this is \emph{not} that the curve is upward-sloping, even though it ususally is. It is that the constract is by construction a no-knockout forward: by buying forward protection today for the period $[t_E,T]$ one is \emph{not} giving up the payout arising from defaults in the time interval $[0,t_E]$. In essence one is buying the same protection but paying for it over a shorter period of time, necessitating a higher running premium.

To find the forward spread from (\ref{eq:sfwd}), the simplest route is iteration, which is effective because the RPV01 does not depend strongly on the spread.
It is also possible to adapt (\ref{eq:newRPV},\ref{eq:invRPV}), noting that the RPV01 is over $[t_E,T]$ rather than $[0,T]$.

Option dealers often quote the forward spread, but they do not always do so and even when they do the results are inconsistent between dealers; then an adjustment can be made to the model option prices to accommodate this if needed.

\subsection{Pricing model, `simple' construction}

The basic idea is to regard the PV of the protection leg, which is the main source of uncertainty, as lognormal. Roughly, this corresponds to a `lognormal spread model', but modelling the PV rather than the spread\footnote{Note that in the context of the single names we avoided this approach for the part-upfront options, because of inconsistency with the all-running case: with the index options there is only one product to model, so we can deal with that as the \emph{de facto} standard.}
. In fact, it is  identical in principle to lognormal interest-rate swaption pricing. In more detail, we are going to write $X$ for the PV of the protection leg of the underlying swap and $Y$ for the PV of the coupon leg at the prescribed strike, and then the option is to swap $X$ for $Y$, which is done using Black--Scholes--Margrabe, where the volatility in question is that of $X/Y$.

Incidentally, a fundamental misapprehension about BSM, which can cause much confusion, is the notion that $X,Y$ need to be jointly lognormal: in fact, it is only their ratio that has to be lognormal, though we need to be careful over the numeraire\footnote{Specifically either that of $X_t$ or $Y_t$, the two being equivalent conditions. The usual derivation, and indeed Margrabe's \cite{Margrabe78}, or  \cite{Bjork98}, Example~19.9, pp.284--5, is done via Black--Scholes assuming $X_t,Y_t$ to be geometric Brownian motions. This is a rather roundabout approach: a more general development is to write down the dynamics of $X_t/Y_t$ directly. To see how this is done, refer to \cite{Martin23a}.}.
We do, however, need both to be positive. Therefore we have to be a little careful in grouping the terms, and the following construction achieves this in an intuitive way:
\begin{eqnarray}
X_t &=& L_t + \frac{N_t}{N_{00}} (s_t-c) \ISDARPV_t(s_t;t,T) + \frac{cN_0}{N_{00}} \extra_t \label{eq:Xt1} \\ 
Y_t &=& \frac{N_0}{N_{00}} \Big( (s_K-c) \ISDARPV_t(s_K;t_E,T) + c \extra_t \label{eq:Yt1} \Big) \\
\extra_t &=&  \ISDARPV_t(s_0;t,T) 
 \label{eq:Wt1} 
\end{eqnarray}
The only source of volatility in $Y_t$ is interest rates, which in context are only of secondary importance: there is no direct or indirect dependence on $s_t$.
The extra term $A_t$ is explained as follows. The expression $X_t$ is the PV at time $t$ of a long index protection trade entered into at time 0 for an investor who is assumed to own a risky annuity of maturity $T$ to pay the cost of carry during the life of the trade, thereby making the trade fully-funded. This is because the first term in (\ref{eq:Xt1}) is the accrued payout from defaults, the second is the PV of the remaining index protection, and the third ($A_t$) is the PV of the remaining cashflows of the annuity.

It is easy to see that $X_t>0$ a.s., because the first term is nonnegative and $\extra_t$ exceeds the negative part of the second term in $X_t$. In fact $X_t$ approaches $0$ in the situation where $N_t=N_{0}$ (so $L_t=0$) and $s_t\to0$.
With $Y_t$ is is not so easy because if $s_K$ is low enough then it will go negative, but this is not thought to be too much of a problem in practice. An alternative construction makes $A_t$ a riskfree annuity, and this avoids the problem, though the results are not quite as neat.
Finally, $X_t$ and $Y_t$ are well-defined even in the Armageddon event, for although $s_t$ and hence $\ISDARPV_t(s_t;t,T)$ are no longer defined, the term that references them is being multiplied by $N_t/N_{00}$ which is zero and so it vanishes.

The discounted expectations of $X_{t_E}$ and $Y_{t_E}$ at time $0$ (today) will be called $\tilde{X_0}$, $\tilde{Y}_0$.   
We deal with $X$ first: by construction
\begin{equation}
\tilde{X}_0 = \frac{N_0}{N_{00}}  s_0\ISDARPV_0(s_0;0,T) 
\label{eq:X0}
\end{equation}
as it is the cost of all-upfront protection from 0 to $T$; given our discussion earlier on forwards, this seems a natural result.

Now for $Y_t$, the first term, i.e.\ the one with $s_K \ISDARPV_t(s_K;t_E,T)$ in it, requires a little care. 
It is understood as the expected `CDSW' calculation on date $t_E$, using $s_K$ as input and with the prevailing rate curve, as seen today. Given the interpretation as a risky annuity, it seems reasonable to write it as $\ISDARPV(s_K;0,T)-\ISDARPV(s_K;0,t_E)$. The problem is that the latter calculation implicitly allows the annuity to have suffered write-downs (defaults) in the interval $[0,t_E]$, but the $\ISDARPV(s_K;t_E,T)$ term in $Y$ is being multiplied by $N_0$, not $N_t$.
A little thought shows that we should handle it as follows: 
use the forward discount-factor curve, obtained by taking the riskfree discount factors $B^\circ(T_j)$ and dividing by $B^\circ(t_E)$;  then apply the spread $s_K$ to it (hazard rate in effect $s_K/(1-\recov)$) to determine the RPV01 in the usual way, and finally multiply by $B^\circ(t_E)$. If, on the other hand, we want to get a direct answer using (\ref{eq:newRPV}), then we take the forward riskfree PV01, which is
\[
\Pi^\circ_{\textrm{fwd}}(t_E,T) =
\Pi^\circ(T)-\Pi^\circ(t_E),
\]
divide by $B^\circ(t_E)$ so that it is actualised at time $t_E$, then `riskify' it using (\ref{eq:newRPV})\footnote{In other words, use $\Pi^\circ_{\textrm{fwd}}(t_E,T)/B^\circ(t_E)$ in place of $\Pi^\circ(T)$ throughout.} and a spread of $s_K$, and finally discount back to today by multiplying by $B^\circ(t_E)$. The final step is done using the riskfree discount factor because the calculation is unaffected by defaults occurring in $[0,t_E]$. Calling the result $\ISDARPV_{\textrm{fwd}}(s_K;t_E,T)$, we then have
\begin{equation}
\tilde{Y}_0 \approx \frac{N_0}{N_{00}} \Big( s_K \ISDARPV_{\textrm{fwd}}(s_K;t_E,T) + c \big( \ISDARPV_0(s_0;0,T) - \ISDARPV_0(s_0^*;0,t_E) - \ISDARPV_{\textrm{fwd}}(s_K;t_E,T) \big) \Big).
\label{eq:Y0}
\end{equation}
The remaining terms involving $A_t$ simply reflect that the (discounted) forward value of a risky annuity is the difference in PV between the two dates.
For ATMF options it is only the first term that matters, which is not surprising.

We now invoke the Black--Scholes--Margrabe formula for valuing the option to exchange one asset for another:
\begin{eqnarray}
C^{+\textrm{idx}}_0 &=& \tilde{X}_0 \Phi(d_+) - \tilde{Y}_0 \Phi(d_-), \\
C^{-\textrm{idx}}_0 &=& \tilde{Y}_0 \Phi(-d_-) - \tilde{X}_0 \Phi(-d_+) \nonumber, 
\end{eqnarray}
\begin{equation}
\quad d_\pm = \frac{\ln(\tilde{X}_0/\tilde{Y}_0)\pm\half\sigma^2(t_E)}{\sigma\sqrt{t_E}} 
\end{equation}
where $\sigma$ is the volatility of $X_t/Y_t$, which in context is the spread volatility.
The delta is given, as usual, by $\Phi(d_+)$, and the so-called strike-delta is given by $\Phi(d_-)$. The gamma is ususally quoted as $\phi(d_+)/(\sigma\sqrt{T}\,s_0$).
As we pointed out, the method fails for very low strike receivers as then $\tilde{Y}_0$ is negative, but we can simply set $d_\pm=1$ and treat the receiver as worthless. 

The aforementioned forward adjustment can be performed by perturbing $\tilde{X}_0$ so as to make $\tilde{X}_0=\tilde{Y}_0$ for an ATMF option.

\subsection{Put-call parity}

By put-call parity have
\[
C^{+\textrm{idx}}_0 - C^{-\textrm{idx}}_0 = \tilde{X}_0 - \tilde{Y}_0.
\] 
When the strike is equal to the forward, this is exactly zero, as is easily seen from (\ref{eq:sfwd}).
However, it is not entirely clear that the market entirely respects this, and there seems to be a slight adjustment in the sense that $C^{+\textrm{idx}}_0 - C^{-\textrm{idx}}_0 $ is slightly positive when $s_K=s_F$. Most likely, this is due to a slight inconsistency in the calculation of the forward\footnote{It may be related to the treatment of the term $c\Pi_0(s^*;0,t_E)$, and more precisely, the difference between that using $s^*=s_0$ and $s^*=0$. A good approximation to that difference is $ch\,t_E^2/2$, where $h$ is the hazard rate, and so it is more important for the longer-dated options (note the dependence is $\propto t_E^2$ not $ \propto t_E$).}, because this effect should not occur.

Another useful check is the case in which the spread is significantly less than the coupon and we price a receiver with $s_K=c$. This case is easy to analyse because there is no upfront payment upon exercise, i.e.\ the `strike term' is zero. Let us reduce $\sigma$ so that the payer is next to worthless. In that case, buying a receiver is no different from selling index protection now, so its price must be the index upfront (received, as $s_0<c$ by assumption) less the expected carry over the period $[0,t_E]$.
It is easily seen that the various terms in (\ref{eq:X0},\ref{eq:Y0}) achieve exactly that.

\subsection{Price-quoted options}

Incidentally the CDX.HY (high-yield) index is quoted as a bond price, but still trades as a swap. (For example, a price of 97.625 means that a buyer of index protection pays $2.375$\% upfront plus 500bp running.) It is convenient to work with the prevailing upfront $U_t$, which pertains to the version of the index at time $t$ (as we said earlier: with names that defaulted before time $t$ removed, because $L_t$ keeps track of the defaulted ones) and an upfront strike $u_K$; these will be negative if the associated bond price exceeds 100. The index payer, which is usually referred to as a put, has value
\[
\ex_t \left[ \displaystyle \frac{\mma_t}{\mma_{t_E}} \left(L_{t_E} + \frac{N_{t_E}}{N_{00}}U_{t_E} - \frac{N_0}{N_{00}}u_K \right)^+ \right].
\]
Our construction is now the same as before, only simpler:
\begin{eqnarray}
\tilde{X}_0 &=& \frac{N_0}{N_{00}} \left( u_0 + c \ISDARPV(s_0;0,T) \right) \\
\tilde{Y}_0 &=& \frac{N_0}{N_{00}} \left(u_K + c \big(\ISDARPV(s_0;0,T)-\Pi^*(s_0;0,t_E)\big) \right) \nonumber.
\end{eqnarray}
The forward upfront is
\[
u_F = u_0 + c\ISDARPV_0(s_0;0,t_E).
\]

\subsection{Example}

Table \ref{table:tab1} shows the option payer and receiver prices for various indices in later December 2025, as described.

\noindent
\begin{table}[!h]

\hspace{-5mm}
\begin{tabular}{rl}
\begin{tabular}[t]{|rrrrrr|}
\hline
Strike & Pay &	Rec &	$\sigma$(\%) &	$\Delta_P$ &	$\Gamma$ (\%) \\
\hline
45 &	43.0 &	2.1 &	33.0 &	0.893 &	2.25 \\
50 &	26.4 &	8.3 &	35.2 &	0.707 &	3.93 \\
55 &	17.6 &	22.2 &	42.2 &	0.504 &	3.80 \\
60 &	12.9 &	40.0 &	49.0 &	0.370 &	3.10 \\
65 &	10.2 &	59.8 &	55.5 &	0.287 &	2.47 \\
70 &	8.2 &	80.3 &	60.7 &	0.229 &	2.01 \\
75 &	6.6 &	101.0 &	64.7 &	0.185 &	1.66 \\
80 &	5.7 &	122.3 &	69.0 &	0.156 &	1.39 \\
85 &	4.7 &	143.5 &	72.0 &	0.130 &	1.18 \\
90 &	4.0 &	165.0 &	75.0 &	0.110 &	1.01 \\
95 &	3.5 &	186.3 &	77.5 &	0.094 &	0.87 \\
100 &	3.1 &	207.8 &	80.5 &	0.084 &	0.77 \\
\hline
\end{tabular} &
\begin{tabular}[t]{|rrrrrr|}
\hline
Strike & Pay &	Rec &	$\sigma$(\%) &	$\Delta_P$ &	$\Gamma$(\%) \\
\hline
45 &	39.6 &	2.0 &	31.5 &	0.890 &	 2.44 \\
50 &	25.0 &	9.4 &	36.7 &	0.684 &	3.95 \\
55 &	17.3 &	23.4 &	44.4 &	0.494 &	3.67 \\
60 &	12.8 &	40.7 &	50.8 &	0.369 &	3.03 \\
65 &	10.0 &	59.5 &	56.6 &	0.286 &	2.45 \\
70 &	8.0 &	61.5 &	61.5 &	0.229 &	2.01 \\
75 &	6.7 &	99.3 &	66.0 &	0.188 &	1.67 \\
80 &	5.7 &	119.7 &	70.0 &	0.158 &	1.41 \\
85 &	4.9 &	140.2 &	73.5 &	0.134 &	1.20 \\
90 &	4.1 &	160.7 &	76.0 &	0.113 &	1.03 \\
95 &	3.5 &	181.3 &	78.5 &	0.097 &	0.89 \\
100 &	3.3 &	202.2 &	82.5 &	0.090 &	0.80 \\
\hline
\end{tabular}

\\
\begin{tabular}[t]{|rrrrrr|}
\hline
Strike & Pay &	Rec &	$\sigma$(\%) &	$\Delta_P$ &	$\Gamma$(\%) \\
\hline
225 &	178.2 &	11.7 &	33.0 &	0.872 &	0.53 \\
250 &	103.8 &	51.6 &	37.1 &	0.643 &	0.84 \\
262.5 &	81.9 &	85.9 &	40.5 &	0.531 &	0.82 \\
275 &	68.8 &	129 &	45.1 &	0.444 &	0.73 \\
300 &	50.0 &	220 &	52.0 &	0.323 &	0.58 \\
325 &	39.4 &	318 &	58.3 &	0.250 &	0.46 \\
350 &	31.2 &	415 &	62.9 &	0.197 &	0.37 \\
375 &	25.5 &	513 &	67.0 &	0.160 &	0.30 \\
400 &	21.1 &	611 &	70.5 &	0.133 &	0.26 \\
425 &	17.2 &	707 &	73.0 &	0.110 &	0.22 \\
450 &	14.7 &	802 &	75.5 &	0.093 &	0.18 \\
475 &	12.8 &	896 &	78.0 &	0.081 &	0.16 \\
500 &	11.4 &	989 &	81.0 &	0.073 &	0.14 \\
\hline
\end{tabular}

 &
\begin{tabular}[t]{|rrrrrr|}
\hline
Strike & Pay &	Rec &	$\sigma$(\%) &	$\Delta_P$ &	$\Gamma$(\%) \\
\hline
$-8$ &	175.0 &	19.5 &	30.0 &	0.825 &	0.55	\\
$-7.5$ & 142.0 & 36.7 &	32.0 &	0.730 &	0.67	\\	
$-7$ &	117.0 &	62.2 &	34.8 &	0.632 &	0.70	 \\
$-6.5$ & 98.1 &	93.0 &	37.5 &	0.545 &	0.68	 \\
$-6$ &	82.9 &	128 &	40.0 &	0.470 &	0.64	 \\
$-5$ &	60.9 &	206 &	44.5 &	0.354 &	0.54	 \\
$-4$ &	47.0 &	292 &	48.8 &	0.274 &	0.44	 \\
$-3$ &	38.0 &	383 &	53.0 &	0.220 &	0.36	 \\
$-2$ &	31.0 &	476 &	56.5 &	0.178 &	0.30	 \\
$-1$ &	26.1 &	571 &	60.0 &	0.149 &	0.25 \\
$+0$ &   20.2 &	665 &	61.5 &	0.118 &	0.21	\\ 
$+1$ &	18.2 &	763 &	65.0 &	0.104 &	0.18 \\
$+2$ &	16.2 &	861 &	68.0 &	0.091 &	0.16 \\
\hline
\end{tabular}

\end{tabular}

\caption{Option prices as of 22-Dec-25, option expiry 18-Mar-26, CDS maturity 20-Dec-30.
Pay and rec prices in bp upfront; strikes in bp except for HY. 
Top left: MA44 ($s_0=51.25$bp; $s_F=55$bp, $\Pi^\circ=4.785$).
Top right: IG45 ($s_0=50.5$bp; $s_F=54.5$bp, $\Pi^\circ=4.645$).
Bottom left: XO44 ($s_0=246$bp; $s_F=261.55$bp, $\Pi^\circ=4.785$).
Bottom right: HY45 ($P_0=107.57$ or 318bp;  $\Pi^\circ=4.645$).
}

\label{table:tab1}
\end{table}

\subsection{Lognormality of spread vs PV}

We have alluded to this above and make some more observations now. First we provide more detail on the connection with IR swaptions.

The payoff of an IR swaption is $(s_t-s_K)\Pi(s_t)$ where $s_t$ is the swap rate and $\Pi$ the PV01. There are two ways to value this under `lognormal' assumptions: (i) change numeraire to $\Pi$ and assume $s_t$ to be lognormal in the $\Pi$-measure; (ii) use Black--Scholes--Margrabe where $X$, $Y$ are the floating and fixed leg PVs. \emph{These give the same result.} There may not seem to be a difficulty here, but watch what happens if one assumes that BSM requires $X,Y$ to be lognormal. The deduction is that (i) is assuming lognormality of par swap rate in the $\Pi$-measure, and (ii) is assuming lognormality of PV, which cannot both be correct, so how can they give the same answer? The point is, of course, that BSM only requires that $X/Y$ be lognormal (in the measure with either $X$ or $Y$ as numeraire), but $X/Y$ is just the par swap rate, so we are back to (i).

Now the difference with the CDS index option is that the payoff is more complicated because the RPV01s are different on the two legs: the `spot leg' depends on the prevailing credit spread, while the `strike leg' does not; both depend on interest rates, but not much. We are therefore left with two different choices: (i) lognormal spread, (ii) lognormal PV (of the protection leg). In (i) we have to do some numerical integration, because $\ISDARPV(s)$ depends on $s$, whereas in (ii), which is the approach described here, we do not, and we can directly use the familiar Black formulae. Note that (ii) cannot theoretically be correct, because the PV cannot exceed unity, whereas the spread is unbounded. Of course, the market does not believe the spread to be lognormal either, which brings us to our next point.

A difference between (i) and (ii) is seen in the implied volatilities: (ii) requires a higher ATMF vol than (i) to match a given ATMF option price, but a lower OTM payer implied vol. In other words, the vol skew is less pronounced if we use (ii), essentially because lognormality of PV is a `fatter-tailed assumption'. The effect is more pronounced in XO than in Main.

Ultimately, our view is that given that a lognormal model `does not work', for the obvious reason that a volatility surface needs to be created to value the options, one may as well use the simplest formulation (closed-form Black) rather than performing an integral over the payoff in the way that Pedersen does. The method is then easily extended to more complex models such as the Merton jump-diffusion, variance Gamma and other variants.

An intriguing `fact' about OTM payers is that the delta and price are almost exactly proportional as $s_K\to\infty$. This is clear in the pricing examples shown, and also in the 2017 examples in earlier editions of this paper. This can be interpreted in two ways. One is that it tells us about the way that vol surfaces are generated by dealers in response to supply and demand. The other is that it tells us that the implied spread distribution has a power-law tail rather than lognormal, and this in turn has implications about implied spread dynamics. Work on this is in progress.

\notthis{
\subsection{Other notes}

We have deliberately left the relevant quantities as PV's rather than converting them back to spreads. One could try to write $X_t$ and $Y_t$ in terms of a corrected forward spread and a corrected strike (the corrections being for the coupon effect and the accrued loss), viz:
\begin{eqnarray*}
\tilde{X}_t &=& s^*_{F,t}\ISDARPV_t(s_t;t,T), \qquad s^*_{F,t} = \frac{N_t}{N_{00}}s_0 + \frac{L_t}{\ISDARPV_t(s_t;t,T)}+ c(\ldots) \\
\tilde{Y}_t &=& s_K^* \ISDARPV_t(s_K;t_E,T), \qquad s_K^* = s_K + c(\ldots).
\end{eqnarray*}
But this does not seem to achieve anything useful: the PV01s are different, and it also unearths the previously-buried problem of what happens in the Armageddon event ($\tilde{X}_t$ is well-defined but $s^*_{F,t}$ is not). Notice incidentally that some of the adjusted forward corrections in the literature\footnote{e.g.\ O'Kane \cite[\S11.7, Eq.~11.9]{OKane08}} can result in a negative adjusted forward and/or strike spread, in the admittedly unlikely situation of the strike being very low. 
} 

\section{Conclusions}

We have shown how to value, in the Black framework, single-name CDS options in which the strike is quoted wholly or partially upfront and shown that the proportion of upfront has a significant impact on price, if consistency with the standard `all-running' case is to be achieved.
We have also pointed out that no-knockout single-name options that are quoted wholly or partially upfront  contain an embedded recovery option, and given a simple explicit formula for the price.
The treatment of index options that we have given here is not revolutionary, but rather is intended to deal with the payout more carefully and intuitively.

\vspace{5mm}

\emph{I acknowledge helpful discussions with Philipp Sch\"onbucher, Ismail Iyigunler, Danny White, Peter Lind, Jesper Andreasen, and the credit quant/trading desks at JP~Morgan, Deutsche Bank and Citigroup.
I also thank  Huong Vu for her kind assistance with the CDS index numerics in 2017.
Email: {\tt richard.martin1@imperial.ac.uk}
}

\clearpage
\appendix

\section{Recovery and default}
 
We discuss the correlation of recovery and default in a multivariate setting (see e.g.~\cite{Tasche04}). This goes well beyond the scope of this paper and is only intended as a brief justification of our previous assertion that the univariate model we have given can be extended to the multivariate case (in infinitely many ways, in fact).

This is most conveniently done with the assistance of a risk-factor, i.e.\ a random variable $\mathcal{A}$ say (which need not be univariate).
Conditionally on $\mathcal{A}$, all recoveries and defaults are completely independent. 
Let a particular credit have a conditional default probability $p(\mathcal{A})$, so that $\ex[p(\mathcal{A})]=\pbar$ (the average default probability, in this case obtained from the CDS market).
Let $F$ be an arbitrary (cumulative) distribution function, and $\beta$ be a coefficient that will couple the recovery variable $Z$ to the risk-factor $\mathcal{A}$.
Define
\[
F_\sharp(z) := \frac{1}{\pbar}\, \ex[F(z-\beta\cdot \mathcal{A} ) p(\mathcal{A})].
\]
Then define the conditional distribution of $Z$ on $\mathcal{A}$ to be 
\[
\pr[Z<z\cdl \mathcal{A}] = F\big(F_\sharp\inv(\Phi(z))-\beta\cdot \mathcal{A} \big).
\]
The distribution of $Z$ conditionally on default of the issuer within the horizon $t_E$, but unconditionally on $\mathcal{A}$, is 
\[
\pr[Z<z \cdl \tau<t_E] = \frac{1}{\pbar}\, \ex \big[ \pr[Z<z\,\cap\tau<t_E\cdl \mathcal{A}] \big]
\]
\[
= \frac{1}{\pbar} \, \ex\left[  p(\mathcal{A}) F\big(F_\sharp\inv(\Phi(z))-\beta\cdot \mathcal{A}\big) \right] = F_\sharp\big(F_\sharp\inv(\Phi(z))\big) = \Phi(z)
\]
which is what we wanted it to be. This procedure of constructing what might be described as `additive factor-copulas' is quite generally applicable and shows that one can always match a given marginal distribution. As stated above, $\beta$ plays the role of a correlation parameter, while the distribution chosen for $F$ is completely arbitrary\footnote{As long as it is continuous. The Gaussian gives, of course, the Gaussian copula.}.

\clearpage

\bibliographystyle{plain}
\bibliography{}

\end{document}